\documentclass[conference]{IEEEtran}
\makeatletter
\def\ps@headings{%
\def\@oddhead{\mbox{}\scriptsize\rightmark \hfil \thepage}%
\def\@evenhead{\scriptsize\thepage \hfil \leftmark\mbox{}}%
\def\@oddfoot{}%
\def\@evenfoot{}}
\makeatother
\usepackage[letterpaper,top=0.75in,bottom=1in,left=0.625in,right=0.625in]{geometry}
\IEEEoverridecommandlockouts

\usepackage{tikz}
\usepackage{cite}
\usepackage{amsmath,amssymb,amsfonts}
\usepackage{algorithmic}
\usepackage{graphicx}
\usepackage{textcomp}
\usepackage{multirow}
\usepackage{xcolor}

\newcommand{\etc} {\emph{etc.}}
\newcommand{\etal}{\emph{et~al. \/}}
\newcommand{\eg}  {\emph{e.g., \/}}
\newcommand{\ie}  {\emph{i.e., \/}}   

\begin{document}

\date{}

\title{HeATed Alert Triage (HeAT): Transferrable Learning to Extract Multistage Attack Campaigns}

\author{\IEEEauthorblockN{Stephen Moskal, Shanchieh Jay Yang}
\IEEEauthorblockA{
\textit{Rochester Institute of Technology}, Rochester, NY, USA
}}

\maketitle

\begin{abstract}
With growing sophistication and volume of cyber attacks combined with complex network structures, it is becoming extremely difficult for security analysts to corroborate evidences to identify multistage campaigns on their network. 
This work develops HeAT (Heated Alert Triage): given a critical indicator of compromise (IoC), \eg a severe IDS alert, HeAT produces a HeATed Attack Campaign (HAC) depicting the multistage activities that led up to the critical event.
We define the concept of ``Alert Episode Heat" to represent the analysts opinion of how much an event contributes to the attack campaign of the critical IoC given their knowledge of the network and security expertise.
Leveraging a network-agnostic feature set, HeAT learns the essence of analyst's assessment of ``HeAT" for a small set of IoC's, and applies the learned model to extract insightful attack campaigns for IoC's not seen before, even across networks by transferring what have been learned.
We demonstrate the capabilities of HeAT with data collected in Collegiate Penetration Testing Competition (CPTC) and through collaboration with a real-world SOC. 
We developed HeAT-Gain metrics to demonstrate how analysts may assess and benefit from the extracted attack campaigns in comparison to common practices where IP addresses are used to corroborate evidences. 
Our results demonstrates the practical uses of HeAT by finding campaigns that span across diverse attack stages, remove a significant volume of irrelevant alerts, and achieve coherency to the analyst's original assessments.
\end{abstract}


\vspace*{-12pt}
\section{Introduction}
Threats of sophisticated and highly impactful cyber attacks have become so common that many organizations have implemented ``Security Operations Centers" (SOC) to investigate, respond to, and hunt potential threats within networks.
SOC's typically implement a tiered structure where a tier 1 analyst triages the network for critical events which may be escalated to a tier 2 analyst who will respond to the incident.
Assume the role of a tier 1 SOC analyst and you observe a critical alert, ``\emph{GPL EXPLOIT CodeRed v2 root.exe access}", targeting a customer database.
While occurring on a critical asset, a single alert may not be enough evidence to escalate to the tier 2 analyst and you must now look for other ``Indicators of Compromise" (IoC) to develop more evidence that the alert was indeed caused by an adversary.
This is known as a ``triage" and is typically a time consuming and mostly manual process sometimes involving multiple analysts to comb through lengthy log files to find other IoC's related to the initial IoC.
With the inflation of network sizes and the general increase of foreign threats broadly targeting any type of organization, many SOC analysts are overwhelmed with the amount of log data from Intrusion Detection Systems (IDS) which hampers their ability to quickly assess their network for threats.

Given a critical alert (an IoC) and IDS alert logs, we explore if machine learning techniques can aid the analyst in the triage process and automatically reveal other steps the adversary took to ``arrive" at their goal.
The compilation of the actions detailing each ``stage" of the attack is called an ``attack campaign" which would describe how, when, and where the attacker learned about the network, gained initial access, and then eventually achieving their goal.
Developing this attack campaign from IDS alert logs can be extremely difficult as the analyst must consider for each alert: the network context, related attributes between the alerts, and their own expertise to determine the relationship between the critical alert and prior alerts.
These considerations sometimes leads to subjectivity of the actual contribution of the alert to the attack campaign.
We envision an automated triage system to reflect the analyst's opinion on the types of events that they believe are a part of an attack campaign and ability to apply that ``thinking" to other triages in the future.

We propose a system, HeATed Alert Triage (HeAT), to perform automated triaging of IDS alerts. 
Given a critical IDS alert, HeAT creates a ``HeATed Attack Campaign" (HAC) using a set of network agnostic features and a small set of analyst defined critical alert episode relations.
In the form of aggregated alerts defined as ``Alert Episodes," the HAC's generated by HeAT tells the story of the attacker's progression leading to a critical event.
HeAT estimates the "Alert Episode HeAT'' for each alert episode with respect to the critical alert to describe the episodes contribution to the attack campaign given how the analyst has interpreted HeAT-value previously.
We develop HeAT with reusability and transferability in mind; we use network agnostic features so that HeAT can uncover attack campaigns for other critical alerts, adversaries, or networks.
We envision HeAT to be used by SOC analysts to display the campaign relating to IoCs and quickly determine any further action if needed. 
Note that we demonstrate the methodology and capability of HeAT with one specific IDS, Suricata, in this work, while the network agnostic features are generalizable to treat heterogeneous alerts and event logs.
Through the use of Suricata alerts collected in CPTC \cite{cptc} and via collaboration with a real-world SOC, this work demonstrates HeAT's ability to:
\begin{enumerate}
    \item Leverage a small amount of analyst-labeled data to learn a model that mimics how the analyst corroborates evidence to extract multistage attack campaigns. 
    \item Extract and prioritize meaningful attack campaigns that cover diverse attack stages, remove irrelevant alerts, and achieve coherent inference as the analyst does.
    \item Be transferable and applicable across networks with extremely small amount of additional labeled data. 
\end{enumerate}

We will fist discuss the related efforts in Section \ref{sec:related}. Section \ref{sec_heat:mth} gives the details of the HeAT process. followed by the design of the HeAT Gain metric in Section \ref{sec:metrics}. Sections \ref{sec:datasets} and \ref{sec:results} describe our design of experiments and datasets used, and discussed what HeAT is capable of doing via specific examples. Section \ref{sec:conclusion} concludes our findings.


\section{Related Work}
\label{sec:related}
The most common method to represent attack campaigns is through the use of Cyber Attack Kill Chains \cite{KCLockheed}. Kill chains concisely describe key phases of an attack campaign such as reconnaissance, exploitation, and data exfiltration to achieve some objective. Initial designs of kill chains describe sequential steps an attacker must take to achieve a goal. In reality, however, attack campaigns are multi-stage and cyclical \cite{KCunified}, often repeating the same stage as the attack progresses. In recent years, more general frameworks have been developed such as MITRE ATT\&CK \cite{mitreattk} and the Action-Intent Framework (AIF) \cite{moskal2020framework} to describe discrete attack types. Our work aims at extracting out the various phase in the attack scenario and let the order of the phases be dictated by the data. This work primarily refers to the Action-Intent Framework (AIF) for its simplicity while maintaining a reference to the broader ATT\&CK framework.

The data-driven extraction of attack campaigns has been studied in depth in the form of Attack Graphs (AG), which has the capability to provide insights into \emph{how} attackers can traverse a network.
AG's use network topology and vulnerability assessments to define potential paths through a network an adversary can exploit.
AG works employ techniques such as alert correlation \cite{qin2004discovering,zhu2006alert,wang2006using,wang2016alert}, process-mining \cite{de2018process,chen2020distributed}, and Markovian-based approaches \cite{fredj2015realistic,ghafir2019hidden} to map observables to pre-existing AG's.
However, these approaches require a significant amount of expert knowledge to configure, create attacker scenario templates, and assumes that the vulnerabilities are known \cite{alserhani2016alert}.

Focusing on related approaches that give AG-like insights without intimate knowledge of the network and vulnerabilities, we find significantly less work within academia.
Navarro \etal presents HuMa \cite{navarro2017huma} and OMMA \cite{navarro2018omma} to extract context from logs, vulnerability databases like CVE and CAPEC, and analyist feedback to find malware behaviors. 
Moskal \etal \cite{moskal2018isi} use a suffix-based Markov chain to derive sequences of aggregated alerts based on their alert characteristics called \emph{attack episodes} so that sequences of episodes could be compared.
Landauer \etal \cite{landauer2019framework} extract from cyber threat intelligence (CTI) reports and applies the knowledge to raw log data to report actionable multi-stage scenarios.
Lastly, Nadeem \etal \cite{nadeem2021sage} present SAGE which employs S-PDFA to extract meaningful AG's from only intrusion alerts and without prior expert knowledge.
A common challenge these works faced is the lack of high quality labeled attack scenario data to comprehensively assess, compare, and validate the identified attack strategies. This draws our effort to present use-cases to demonstrate the usability of HeAT by mimicking how analysts may use it with minimal effort to provide labeled data.

In the private sector, where data is more abundant, the concept of AI-driven products to assess and automatically triage a network is an extremely fast growing area.
As of 2021, the adoption of AI/ML techniques to solve cyber security problems has exploded.
To name a few, companies such as DarkTrace with their ``Cyber AI Analyst" \cite{darktrace}, IBM with QRadar Advisor with Watson \cite{ibmwatson}, and Centripetal with AI-Analyst \cite{centripetal} all advertise their capabilities to leverage AI specifically to aid analysts in the triaging process.  
While these products are undoubtedly extremely sophisticated due to their substantial resources, it is impossible to assess their true capabilities due to the proprietary nature of the method and data.
While we do not claim to compete with these products, the competition between these products shows that identifying and understanding attack campaigns is \textit{hard}.
Usage of these algorithms and products often requires immense expertise or hard-to-get data-sets; we develop a process to simplify the labor required from the analyst while able to learn and transfer the learned model to extract and prioritize meaningful multistage attack campaigns.

\section{H{e}ATing Episodes to Extract the H{e}ATed Attack Campaign (HAC)} \label{sec_heat:mth}
\begin{figure}[htp]
    \centering
    \includegraphics[width=8.5cm]{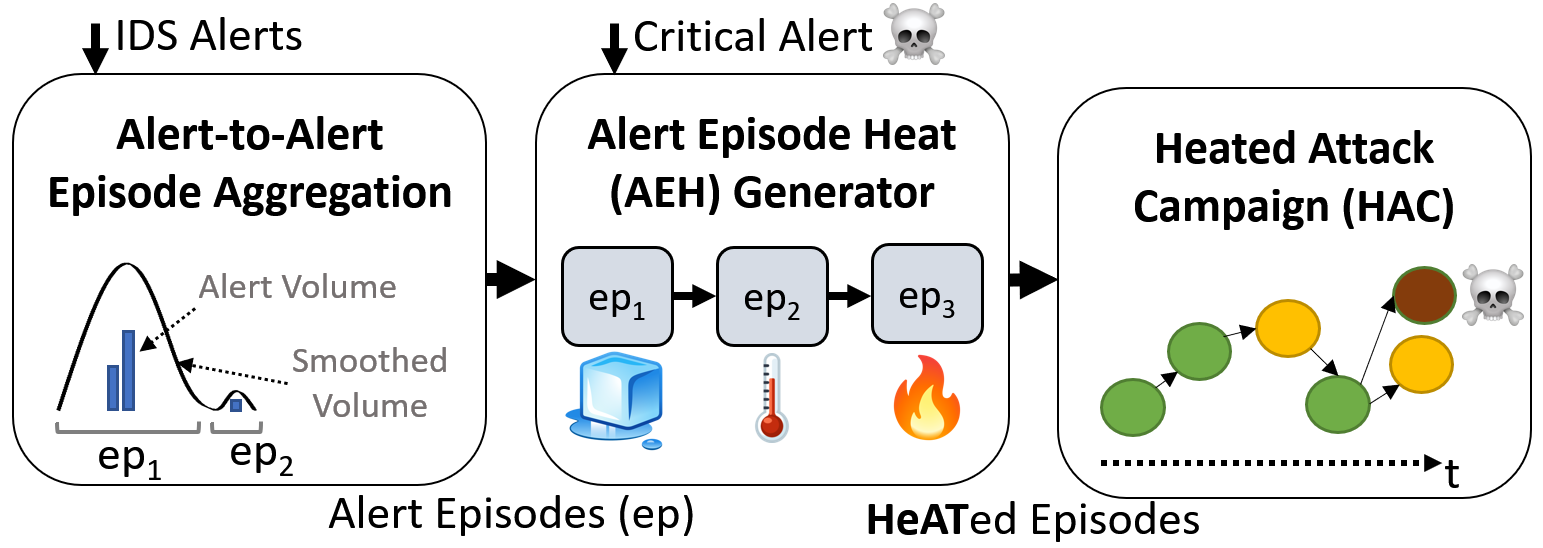}
    \caption{A conceptual overview of the HeAT Process.} 
    \label{fig:heat_overview}
\end{figure}

Given the IDS alert logs from a network and an IoC such as a critical IDS alert, our objective is to identify a sequence of alerts likely to be related to the IoC, forming the attack campaign of the adversary.
We define an ``Attack Campaign" as the collection of actions in time which describe each stage of an attack conducted by an adversary leading to some objective.
As there will be no ground truth describing the real attack campaign, we rely on an initial triage to establish characteristics of an actual attack campaign first.
Then, we address other technical challenges such as high alert volume, network-specific attack characteristics, and limited analyst data-labeling resources to extract meaningful and concise attack campaigns.
The summary of the methods used in HeAT are described below and the system overview is shown in Figure \ref{fig:heat_overview}.

\begin{itemize}
    \item Introduce ``Alert Episode Heat" (HeAT-value) as a numeric ranking system (0-3) representing key milestones of an attack campaign leading towards an IoC.
    \item Use an attack stage-based Gaussian smoothing approach to alert aggregation to create alert episodes indicative of actions performed by adversaries.
    \item Use alert episodes to derive network agnostic features relating characteristics between episodes, enabling prediction of the HeAT-value regardless of attack type or network configuration.
     \item Propose an efficient process to capture the analyst's reflection of meaningful relationships between alert episodes contributing to the same attack campaign.
    \item Use AI/ML to learn and predict HeAT-values for prior episodes to a critical episode.
    \item Construct and visualize HeATed Attack Campaigns (HAC) with ``HeATed episodes''.
    \item Propose the entropy-based HeAT-gain metric to prioritize HAC's with desirable attack campaign properties.
\end{itemize}

\vspace*{-24pt}

\subsection{HeAT Value: Progress Towards Attack Objective} \label{sec:heat}
The concept of Alert Episode Heat, or the HeAT-value, is a numeric ranking system (0-3) which given a critical alert episode $e_c$ and a prior episode $e_p$, the HeAT-value ranks the contribution of $e_p$ to the attack campaign of $e_c$.
We use the concept of ``alert episodes" to represent groups of alerts that are indicative of action(s) with a specific impact.
Each alert episode may contain one or many alerts sharing similar attributes, such as attack impact, which may or may not be related to the campaign of $e_c$.
The HeAT-value is intended to capture the attacker's progression towards $e_c$ given the alerts of $e_p$.

While many IDS's already have some notion of severity embedded within the alert (Suricata's severity attribute), these are typically static and independent from all other alerts that have occurred.
IDS's such as Suricata have no notion of correlated alerts but simply report suspicious behavior based on signature matches of known adversarial actions and additional information is needed to determine if two events are correlated.
Additional factors such as the network topology, the assets contained on specific machines, and the analyst's own expertise is considered when correlating the true severity between security events.
The concept of HeAT is to create these correlations between a critical episode and the episodes prior.
Given $e_c$ and $e_p$, our objective is to define an Heat Generator as: $h(e_p|e_c)=f(e_p,e_c)$ where $\{h \in \mathbb{R}|0 \leq h \leq 3\}$.   

We design the HeAT-values as a small set of discrete values that signify key milestones within an attack campaign.
Table \ref{tab:heat} describes the characteristics of the distinct HeAT-values used to label and create the initial triage training set.
We use the high-level attack stages such as ``reconnaissance", ``exploitation", and ``actions on objective" \cite{KCLockheed} to represent heat levels 1, 2 and 3, respectively, to reflect their progressive impact on a network.
We choose a ``less-is-more" approach as we embed specific attack stage information within our labels and human studies show that 3 to 4 options is optimal reduce error for human surveys \cite{surveycount}.
With HeAT level representing a small number of mutually exclusive attack stages, we believe the analyst can quickly determine an appropriate HeAT level and we believe there will be less ambiguity between HeAT levels.  

\begin{table}[htbp!]
\centering
\caption{Description of the HeAT-values.}
\label{tab:heat}
\begin{tabular}{|c|l|}
\hline
HeAT-value & \multicolumn{1}{c|}{Description} \\ \hline
0          & No relation to critical event    \\ \hline
1 & \begin{tabular}[c]{@{}l@{}}Recon. actions that may provide info. about $e_c$\end{tabular} \\ \hline
2 & \begin{tabular}[c]{@{}l@{}}Exploitation of assets giving access required to achieve $e_c$ \end{tabular} \\ \hline
3 &  \begin{tabular}[c]{@{}l@{}}Exfiltration/DoS/Access to info. directly relevant to $e_c$\end{tabular} \\ \hline
\end{tabular}
\end{table}


\subsection{Alert Episodes with the Action-Intent Framework (AIF)}
A common problem with IDS's is the high volume of alerts presented to the user and that the descriptions contained within the alerts are often difficult to resolve \textit{what} the attacker was attempting to achieve \cite{moskal2020framework}.
This alone makes the triaging process extremely time consuming.  
Alert aggregation is used to group alerts of similar attributes such as time proximity, IP addresses, or effect of the attacks to reduce the number of events presented to the analyst.
One main challenge here is that we only have the attributes defined within the Suricata IDS alerts. Applying PATRL \cite{moskal2021translating}, we map the alert signature description to the attack intent stages (AIS) the adversary could be performing.
Given the AIS, we define an  `Alert Episode' to be the set of aggregated alerts for a single source IP and the same attack stage across multiple target IP's within a close time proximity.

Moskal \etal \cite{moskal2018isi} describes a process where alerts are aggregated based on the fluctuations of alert volume within a time window for specific IP addresses and Suricata categories to uncover common sub-sequences of attack patterns.
They use a Gaussian Smoothing approach to aggregate alerts based on source IP, attack stage, and time can aggregate alerts based on similar impact within a time window.
We choose this process due to its effective application of Gaussian smoothing to represent aggregate alerts whose the alert arrival time may be inconsistent, sporadic, or periodic.
We expand on this Gaussian smoothing process by the adoption of the Action-Intent Stages (AIS) to aggregate signatures based on similar attacker intent \cite{moskal2020framework}. 
With the provided mapping of Suricata description to AIS \cite{moskal2021translating}, we create episodes of alerts that are likely caused by a single adversary with a similar impact.
It is the collection of these alert episodes related to a single objective or goal that describes the overall attack campaign.

To create the episodes, Gaussian low-pass filtering is applied to histograms in time of alert volume for single IP and AIS, where the LPF filter parameter is set based on the expected duration of the action on a per AIS basis.
We accept that source IP is not a reliable attribute to observe the , we believe it is the best opportunity to capture alerts caused by one adversary.
Certain types of attacks may have longer duration than others and thus different filter sizes are used.
Our Alert Episodes are derived by evaluating each peak of the AIS-based filtered histograms and the collection alert(s) contained in-between the two local minima of the corresponding peak make the episode.
Conducting this process over each attack stage for each source IP, combining the derived episodes, and sorting the by the peak episode time gives an abbreviated view of the sequence of ``actions" performed by that adversary.

\subsection{Network Agnostic Features between Alert Episodes}
We engineer our features with two elements in mind: 1) the features describe relations between two episodes so that the HeAT-value can be determined with respect to a critical episode and 2) the features are network agnostic so that the model does not learn network specific HeAT relations that cannot be applied to other attack types or network configurations.
As the episodes contain set of alerts with a wide variety of complex data types such as IP addresses, alert signatures, \etc, we manually define a set of episode features to represent each of these data types.
Each alert episode contains the attributes shown in Table \ref{tab:eps} which are derived from the alerts contained within the episode.

\begin{table}[htbp!] 
\centering
\caption{Definitions of the attributes contained within an alert episode.}
\begin{tabular}{|l|l|l|}
\hline
\multicolumn{1}{|c|}{Name} & \multicolumn{1}{c|}{Symbol} & \multicolumn{1}{c|}{Description} \\ \hline
Ep. Peak                   & $e_{peak}$                  & Time of peak alert volume        \\ \hline
Ep. Start                  & $e_{start}$                 & Time of earliest alert           \\ \hline
Ep. End                    & $e_{end}$                   & Time of latest alert             \\ \hline
Distinct Sources         & $e_{src}$                   & Set of distinct src IP(s)     \\ \hline
Distinct Targets         & $e_{tgt}$                   & Set of distinct tgt IP(s)     \\ \hline
Distinct  Sigs           & $e_{sig}$                   & Set of distinct signatures       \\ \hline
Distinct Tgt. Ports     & $e_{port}$                  & Set of distinct dest. ports      \\ \hline
AIS                        & $e_{ais}$                   & AIS of the episode               \\ \hline
\end{tabular}
\label{tab:eps}
\end{table}

We define three types of features to capture different aspects of common characteristics between episodes: 1) Time, 2) IP, and 3) Action based features, shown in Table \ref{tab:feats}.  
The time-based features capture the differences between the critical alert episode and the prior episodes.  
Our IP-based features compare if there are similarities between IPs of the two episodes without defining any details of the IPs themselves.
Lastly, our action-based features capture similarities between attack stages, signatures, and port numbers to determine if the two episodes have a similar network impact.

\begin{table}[hbtp!]
\centering
\caption{The set of network agnostic features relating the attributes of two alert episodes.}
\begin{tabular}{|c|l|l|}
\hline
Type                    & \multicolumn{1}{c|}{Feature} & \multicolumn{1}{c|}{Description}                                                                      \\ \hline
\multirow{4}{*}{Time}   & Ep. Interval Overlap         & \begin{tabular}[c]{@{}l@{}}Overlap of the start \&\\ end times of $e_c$ and $e_p$\end{tabular} \\ \cline{2-3} 
                        & Ep. Peak Time Diff.          & $e_{c,peak} - e_{p,peak}$                                                                             \\ \cline{2-3} 
                        & Ep. Start Time Diff.         & $e_{c,start} - e_{p,start}$                                                                           \\ \cline{2-3} 
                        & Ep End Time Diff.            & $e_{c,end} - e_{p,end}$                                                                               \\ \hline
\multirow{6}{*}{IP}     & Has Matching Source          & 1 if $e_{c,src} \cap e_{p,src}$ else 0                                                                \\ \cline{2-3} 
                        & Has Matching Target          & 1 if $e_{c,tgt} \cap e_{p,tgt}$ else 0                                                                \\ \cline{2-3} 
                        & Matching Source Ratio        & \begin{tabular}[c]{@{}l@{}}Ratio of matching\\ source IPs \end{tabular}                                                                      \\ \cline{2-3} 
                        & Matching Target Ratio        & \begin{tabular}[c]{@{}l@{}}Ratio of matching \\ target IPs \end{tabular}                                                                           \\ \cline{2-3} 
                        & Crit. Source as Target    & 1 if $e_{c,src} \cap e_{p,tgt}$ else 0                                                                \\ \cline{2-3} 
                        & Crit. Target as Source    & 1 if $e_{c,tgt} \cap e_{p,src}$ else 0                                                                \\ \hline
\multirow{5}{*}{Action} & Critical Ep. AIS             & 1-hot encoded $e_{c,AIS}$                                                                             \\ \cline{2-3} 
                        & Prior Ep. AIS                & 1-hot encoded $e_{p,AIS}$                                                                             \\ \cline{2-3} 
                        & Has Matching Sigs.      & 1 if $e_{c,sig} \cap e_{p,sig}$ else 0                                                                \\ \cline{2-3} 
                        & Matched Sig. Ratio           & Ratio of matching sigs.                                                                          \\ \cline{2-3} 
                        & Matching Dest. Port      & 1 if $e_{c,p} \cap e_{p,p}$ else 0                                                              \\ \hline
\end{tabular}
\label{tab:feats}
\end{table}

Our hypothesis is that these network agnostic features will allow us to uncover a variety of attack campaigns without detailed network topology or system vulnerabilities.
We propose that these network agnostic features can be used to predict the HeAT-value for other attack types and be applied to other networks.
In the next section we describe our methodology for creating the ``Heat Generator" and how we leverage a small amount of labeled HeAT-values to determine HeATed Attack Campaigns (HAC).

\subsection{Heat Generator: Learning the Analyst's Heat} \label{sec:generator}
The ``heat generator" is our name for a machine learning model for predicting the HeAT-value given the aforementioned network agnostic features representing the relationship between two alert episodes. 
When defining the concept of the Heat Generator significant challenges arise as labeled data describing an attack campaign with respect to IDS alerts generated is few and far between.
Data sets that do exist within the research community are typically either outdated (irrelevant attack types), unlabeled, and/or represented in a different domain (\ie packet captures) than IDS alerts.
Instead we have the user conduct an initial ``triage" of their IDS alerts, label HeAT-values to episodes related to a known IoC, and then use the network-agnostic features to create a predictive model to ``generate" heat given other IoC's.

Given an set of traiged episodes with the analyst's labelled HeAT-value, we define the heat generator as $h(e_p|e_c)=f(e_p,e_c)$ where $\{h \in \mathbb{R}|0\leq h \leq 3\}$.
We define a HAC for given a critical episode $e_c$ as the set of all prior episodes $e_p \in E$ where the heat generator applies a non-zero HeAT-value.
Our requirements for a selecting a machine learning model for this application are bound by our non-linear features and that the HeAT-value must be a continuous value.
HeAT is implemented in Python and our heat generator leverages Fast.AI's Tabular learner \cite{fastai} to predict the HeAT-value.
All features within our data are standardized to have a zero mean and unit variance and we report our 5-fold cross-validated mean squared error (MSE) for the training data.   

The process of extracting the HAC from the our data is similar to our training process where a critical alert is given by the user, HeAT finds the corresponding episode containing the critical alert, and then the heat generator is used to ``HeAT" \emph{all} prior episodes with respect to the critical episode.
We apply HeAT to all prior episodes to give our model the opportunity to discover episodes that may have significantly contributed to the attack campaign that may not immediately obvious.
The set of HeATed episodes with a non-zero HeAT-value are then considered to be apart of the HeATed attack campaign of the critical alert. 
As the episodes may contain many alerts, we foresee the generator finding small relations to the critical episode and apply a small amount of HeAT to episodes that may not contribute much to the overall campaign; a minimum HeAT-value threshold can be applied if the user desires.
Truly impactful episodes will have higher HeAT-value levels than those with just a few similarities between features.


\section{Metrics to Assess HAC: H{e}AT-Gain}
\label{sec:metrics}
To demonstrate the usefulness of HeAT in its ability to reveal meaningful attack campaigns to the analyst, we develop a quantitative entropy-based metric called ``HeAT Gain."
Note that the real-world usage of HeAT will not have the true step-by-step campaign laid out to calculate the accuracy of an extracted HAC.
Therefore, we propose using HAC characteristics to infer on the ``usefulness" of the HAC without labeled ground truth data.
We consider the generated HAC, the data being tested (history of IDS alerts), and the HeAT model training data to develop a quantitative metric to evaluate the value the HAC provides to the analysts.
Using HeAT Gain, the users can quickly compare and prioritize HAC's especially in production systems with high volumes of daily traffic.

Given the attack episodes of the data-under-test ($E_d$), the set of heat-labeled training episodes ($E_t$), and an HAC representing a finite set of HeATed attack episodes ($E_h$ where $E_h \subseteq E_d$), we define HeAT Gain, $\Delta(E_h|E_d, E_t)$ as a combination of three entropy-based metrics:
\begin{itemize}
    \item \textbf{AIS Coverage Gain} reflects the breadth and variety of attack stages covered by $E_h$; a higher value indicates more completed coverage of the attack progression by the extracted HAC.
    \item \textbf{Noise Reduction Gain} reflects the reduction of noisy and irrelevant episodes from $E_d$ to $E_h$; a higher value indicates a more significant reduction.
    \item \textbf{HeAT Coherence} reflects how well the HeAT model predictions match the analyst's initial assessments that define the training set $E_t$; the closer to 0 the more coherent the model is to the analyst's opinions.
\end{itemize}


Let $A_x: \{a \in \mathcal{A} \}$ be the random variable of attack stages contained in the set of episodes $E_x$, and $Y_x: \{y \in \mathbb{Z} ~ | ~ 0 \leq y \leq 3\}$ the random variable of the predicted HeAT values of these episodes. We calculate Shannon's entropy, as shown below, with respect to $A_x$ with the log base $b=|\mathcal{A}|$. The subscript $x$ is a place holder for $h$, $d$, and $t$ denoting the HAC,  data-under-test, and training data, respectively.

\begin{equation}
    H(A_x)=-\sum_{a\in\mathcal{A}}P(a)log_b P(a)
\label{eq:entropy}
\end{equation}

\begin{equation}
    H(A_x|Y_x)=-\sum_{\substack{a\in\mathcal{A} \\ y=0,1,2,3}}P(a, y)log_b \frac{p(a, y)}{p(y)}
\label{eq:centropy}
\end{equation}

\textbf{AIS Coverage Gain: $\delta_{ACG}=H(A_h)$}.  
AIS Coverage Gain quantifies the diversity of attack stages captured in a HAC. HAC that covers a diverse set of attack stages signifies the capturing of evidences reflect each step of the attack campaign progression, from reconnaissance to achieving the final attack objective. A low coverage gain may indicate limited usefulness of the extracted HAC due to, \eg a specific attack action or large volume of scanning that do not lead to critical attacks.

\textbf{Noise Reduction Gain: $\delta_{NRG}=H(A_d)-H(A^{'}_h)$}.
The Noise Reduction Gain measures the reduction in AIS coverage randomness achieved by HAC. Note that while we want to achieve diverse coverage of attack stages, HAC is meant to filter out episodes that are not relevant to the IoC -- recall the HeAT threshold. Define $A^{'}_{h}: \{A_h \cup \alpha_r\}$, $\alpha_r$ represents all irrelevant episodes that were filtered out from $E_d$ by the HAC. A useful HAC will remove a large number of irrelevant episodes, making $H(A^{'}_h)$ smaller than $H(A_d)$ and thus a larger Noise Reduction Gain.
This metric complements the AIS Coverage Gain in that they together reward HAC's that contain diverse attack stages ($A_h$) yet reduced a significant amount of irrelevant episodes ($\alpha_r$).

\textbf{HeAT Coherence: $\delta_{COH}=abs(H(A_h|Y_h)-H(A_t|Y_t))$}.
HeAT Coherence measures the difference in ``how HeAT values infer attack stages'' between how the model does for the HAC and how the analyst believe such relationship should be. Note that HAC is meant to aid the analyst by mirroring her/his analytical thinking on how an attack campaign might transpire. The HeAT values predicted by the machine learned model should resemble well how the analyst might assign the value if they were to do it. In the absence of ground truth labels in the test data (to mimic production systems), we compare the conditional entropy of the model inference versus that of the analyst's assignment in the training data. 
Ideally $H(A_h|Y_h)=H(A_t|Y_t))$. In reality, $H(A_h|Y_h)>H(A_t|Y_t))$ indicates factors other than attack stages giving more influence to the HeAT values than what the analyst has shown; whereas $H(A_h|Y_h) < H(A_t|Y_t)$ indicates that is overly apply HeAT based on the attack stages.
Either of these divergent cases are undesirable and may indicate the model is insufficient and require additional analyst input for training data.

\textbf{HeAT Gain: $\Delta(E_h|E_d,E_t)=\delta_{ACG}+\delta_{NRG}-\delta_{COH}$}.
The overall HeAT Gain metric rewards HAC's that cover a diverse set of attack stages, filter out irrelevant episodes, and are coherent with how analyst assesses the HeAT values. The overall metric and its components will be used to evaluate the usefulness of the extracted HAC's in the next section. Note that in production systems, there are likely to be many HAC's extracted by HeAT given a set of potential IoC's. The HeAT Gain thus can be used to prioritize the extracted HAC's. As we will show in the next section, examining HAC's with high HeAT Gains allows the users to focus on analyzing meaningful campaigns instead of buried in overwhelming alerts.

\section{Design of Experiments and Datasets}
\label{sec:datasets}

To demonstrate the effectiveness and usefulness of HeAT in a practical setting, we consider two sets of Suricata alerts collected through 1) CPTC and 2) collaboration with a real-world SOC.
CPTC data is considered because there are abundant attack activities through student penetration tests in a controlled environment. Such activities allows our research team to curate the initial training data with HeAT values associated with multistage attacks.
Using the small set of curated labeled data, we can then assess HeAT for the remaining unknown campaigns for critical IoC's in CPTC.
Assessing HeAT using the remaining CPTC data offers detailed insights on how the extracted campaigns in a controlled environment are useful in terms of the three HeAT-Gain components discussed earlier. 
We then further test how well the learned HeAT model can be applied to a real-world SOC environment where both the adversary behavior and network configurations are not known to our research team.
As we will show in Section \ref{sec:results}, HeAT applies very well using the network agnostic features, and further improves by adding an extremely small set of training data with respect to the SOC environment.

\subsection{CPTC and SOC Datasets}
Table \ref{tab:datastats} summarizes the number of alerts and episodes as well as the unique source and target IPs and unique Suricata alert signatures for the CPTC data. There were 10 teams in this competition and we set aside one team to curate the initial training data through detailed triaging and assigning HeAT values to alert episodes relevant to specific IoC's. The team selected for training contains a good number and diverse range of alerts for us to elicit analyst's opinion of how attack campaigns transpired.

\begin{table}[hbtp!]
\centering
\caption{Statistics of alerts and episodes from all teams in CPTC and the team used for the initial training triage.}
\label{tab:datastats}
\begin{tabular}{|c|l|l|l|l|l|}
\hline
 &
  \multicolumn{1}{c|}{\begin{tabular}[c]{@{}c@{}}Unique\\ Sources\end{tabular}} &
  \multicolumn{1}{c|}{\begin{tabular}[c]{@{}c@{}}Unique\\ Targets\end{tabular}} &
  \multicolumn{1}{c|}{\begin{tabular}[c]{@{}c@{}}Unique\\ Sigs\end{tabular}} &
  \multicolumn{1}{c|}{\begin{tabular}[c]{@{}c@{}}Total\\ Alerts\end{tabular}} &
  \multicolumn{1}{c|}{\begin{tabular}[c]{@{}c@{}}Total\\ Episodes\end{tabular}} \\ \hline
All &
  45 &
  81 &
  265 &
  169,448 &
  3200 \\ \hline
Train & 29 & 49 & 171 & 53,362 & 529 \\ \hline
\end{tabular}
\end{table}

The SOC dataset contains 7 days of Suricata alerts from a medium-sized higher-ed institution.
Table \ref{tab:soc_stats} shows the the alert and episode volumes for each day, along the unique source and target IPs and unique alert signatures for each of the 7 days. 
One may immediately notice the orders of magnitude increase in volume and diversity of IP addresses involved in the SOC data comparing to the CPTC data.
Interestingly, the number of unique signatures are not that much different between the two datasets\footnote{Unique signatures overlap from day to day in the SOC data.}. 
For SOC alerts, there are on average about 35\% of unique alerts captures per day are labeled as `high-severity' as defined by Suricata.
Note, however, that in reality not all of these alerts are critical and could distract the analysts from discovering the truly critical attack campaigns.

\begin{table}[hbtp!]
\centering
\caption{Daily summary of alert traffic for the SOC data-set (*Weekend Traffic) -- highest total are shown in bold.}
\label{tab:soc_stats}
\begin{tabular}{|c|l|l|l|l|l|l|}
\hline
Day &
  \multicolumn{1}{c|}{\begin{tabular}[c]{@{}c@{}}Total\\ Alerts\end{tabular}} &
  \multicolumn{1}{c|}{\begin{tabular}[c]{@{}c@{}}Total\\ Episodes\end{tabular}} &
  \multicolumn{1}{c|}{\begin{tabular}[c]{@{}c@{}}Unique\\ Sources\end{tabular}} &
  \multicolumn{1}{c|}{\begin{tabular}[c]{@{}c@{}}Unique\\ Targets\end{tabular}} &
  \multicolumn{1}{c|}{\begin{tabular}[c]{@{}c@{}}Unique \\ Sigs.\end{tabular}}  \\ \hline
1*  & \textbf{894,915} & 77,202          & 3,071          & \textbf{195,553} & 123                   \\ \hline
2*  & 664,754          & 74,396          & 2,946          & 187,452          & 104                  \\ \hline
3  & 620,202          & 94,408          & 5,263          & 171,006          & 145                  \\ \hline
4  & 773,622          & \textbf{97,276} & 5,521          & 189,877          & 145                 \\ \hline
5 & 532,182          & 89,919          & \textbf{5,357} & 163,818          & \textbf{149}  \\ \hline
6 & 549,498          & 87,929          & 5,072          & 165,265          & 125                   \\ \hline
7  & 646,699          & 83,409          & 4,860          & 186,681          & 130                 \\ \hline
\end{tabular}
\end{table}

Due to the large number of source IPs, we use the Autonomous System Number (ASN) derived from the source IPs to perform our Gaussian-smoothing aggregation (recall Secion 3.2). We expect this to be a realistic and practical approach in real operational environment where external sources are differentiated in a regional instead of per IP basis.
By using ASN, we reduce the number of episodes across the 7-day period from 604,537 to 336,822, a 44.3\% reduction.
Note that with the network-agnostic features, our learning model through the use of IPs in the training data are still applicable when applying ASN-based aggregation.
This further demonstrates how our process' ability to transfer across networks.

\section{HeATed Attack Campaign Analysis} \label{sec:results} 

\subsection{Evaluating HAC's with HeAT Gain}

\subsubsection{CPTC: the use of HeAT gain components} 
Since the CPTC dataset contains multiple teams penetrating the same network, we consider the same critical IoC that is prevalent across teams (or even within the same team but occur multiple times).
Specifically, we consider the alert ``\emph{GPL EXPLOIT CodeRed v2 root.exe access}" -- the third case in our training data.
This will allows us to establish a close-to-best-case scenario to assess how well HeAT performs to extract HAC's for the same type of IoC that appears in the training data.

This CodeRed signature appears 38 times, out of the 169,448 alerts in the CPTC test alerts.
Without HeAT, an analyst would have to manually identify which of the remaining alerts are related to each CodeRed occurrence through numerous SIEM queries.
The application of our alert episode aggregation alone would make this analysis significantly less tedious as the aggregation process produces 3.200 alert episodes.
We use our HeAT Gain metric to prioritize the assessment of attack campaigns of the 38 potential campaigns leading towards `CodeRed' and also compare HeAT against typical methods one would use if performing a manual assessment.
We consider three `naive' rule-based methods for generating attack campaigns for a given IoC by collecting the sets of episodes where: 1) the source IP matches with the IoC, 2) the target IP matches with the IoC, and 3) both the source and target IPs match the IoC.

\begin{figure*}[hbtp!]
    \centering
    \includegraphics[width=18cm]{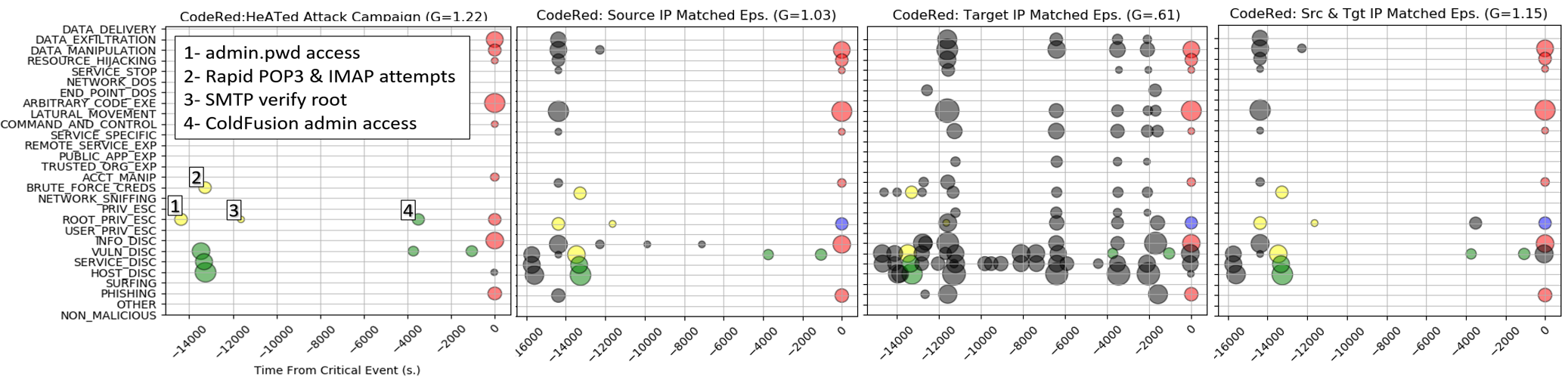}
    \caption{Four HAC's generated for a CodeRed IoC using (from left to right) HeAT, matching src IP, matching tgt IP, and matching src and tgt IPs. Each circle (from grey, green, yellow, to red signifying low to high heat level) represent an alert episode with the size reflecting the number of alerts in the episode.}
    \label{fig:cr_all}
\end{figure*}

Figure \ref{fig:cr_all} compares the CodeRed attack campaigns generated by HeAT versus those generated using the three IP-based baseline methods.
Each circle represents an alert episode for a specific attack stage, with its size reflecting the number of alerts in the episode and its color corresponding to the predicted HeAT value (from grey, green, yellow, to red signifying low to high HeAT).
As can be seen, HeAT gives the highest Heat-Gain (G=1.22) HAC (left), which is visually the most concise attack campaign with explicit attack stages from reconnaissance, exploitation, admin accesses, leading to CodeRed.
In contrast, if one were to inquire alert episodes based on by the IoC's target IP, the resulting HAC (second to the right) will lead to substantially more episodes, many of which are actually not relevant.
The HAC's extracted with the other two baseline methods are similar, better than the target IP based approach does, but still not as useful as the one generated by HeAT.

\begin{figure}[hbtp!]
    \centering
    \includegraphics[width=7.7cm,height=5cm]{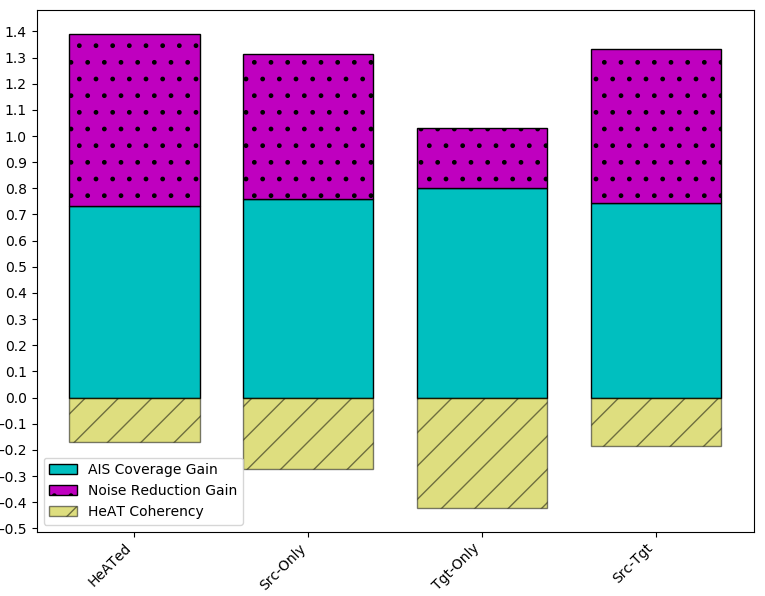}
    \caption{The HeAT Gain components for the HACs generated by HeAT and the three baseline approaches.}
    \label{fig:cr_comp}
\end{figure}

A detailed examination of the HeAT Gain components for each of the four cases give the bar charts shown in Figure \ref{fig:cr_comp}. 
Recall that there are HeAT Gain components: 1) AIS Coverage Gain, 2) Noise Reduction Gain, and 3) HeAT Coherency. The three components measures the usefulness of an HAC in its ability to display diverse attack stages, remove irrelevant episodes, and achieve coherency to the analyst's campaign assessment approach, respectively.
As can be seen, all four HAC's have similar AIS Coverage Gain while the HAC generated by HeAT (bar on the left) shows the most noise reduction and the most coherency. Using both the source and target IP (bar on the right) seems to achieve a similar Gain (1.15) to that achieved by HeAT (1.22). The difference in these two cases lies in the reduction of irrelevant episodes, which is why the HAC generated by HeAT is visually more concise and interpretable as shown in Figure \ref{fig:cr_all}. 


\subsubsection{SOC: assess usability in real-world operation}

With the trained HeAT model using the CPTC data, we hypothesize that the HAC extraction can be sufficiently effective when applying to a different network with the network agnostic features.
A challenge in assessing the real-world SOC data is the significant number of IoCs and alert episodes: $\sim$50K episodes for thousands of potential IoC's per day. Using the HeAT Gain, we mimic how the analyst could do by sorting the HAC's of these potential IoC's using the HeAT gain metric. 
For this paper, we choose 2 Suricata alert signatures as examples to demonstrate the usefulness of HeAT:
`WEB\_SPECIFIC\_APPS Drupalgeddon2\textless{}8.5.1 RCE through Reg. CVE-2018-7600' and
`XPLOIT F5 TMUI RCE vulnerability CVE-2020-5902'.
These two signatures contain CVE numbers, providing a reference of the corresponding vulnerabilities and associated exploits. Each of these signatures appear hundreds of times across the 7-day period of the SOC data. We imagine a use-case of HeAT Gain where the analyst chose a threshold, \eg with respect to the AIS Coverage Gain, to prioritize the top HAC's to focus on. 

The above process is expected to be configurable in the production system of HeAT. In short, the configuration will give 7 HAC's in the 7-dayperiod for each signature if using an AIS Coverage Gain threshold of .4 -- a much more manageable attack campaign analysis than hundreds of possibilities.

\begin{figure*}[hbtp!]
    \centering
    \includegraphics[width=14cm]{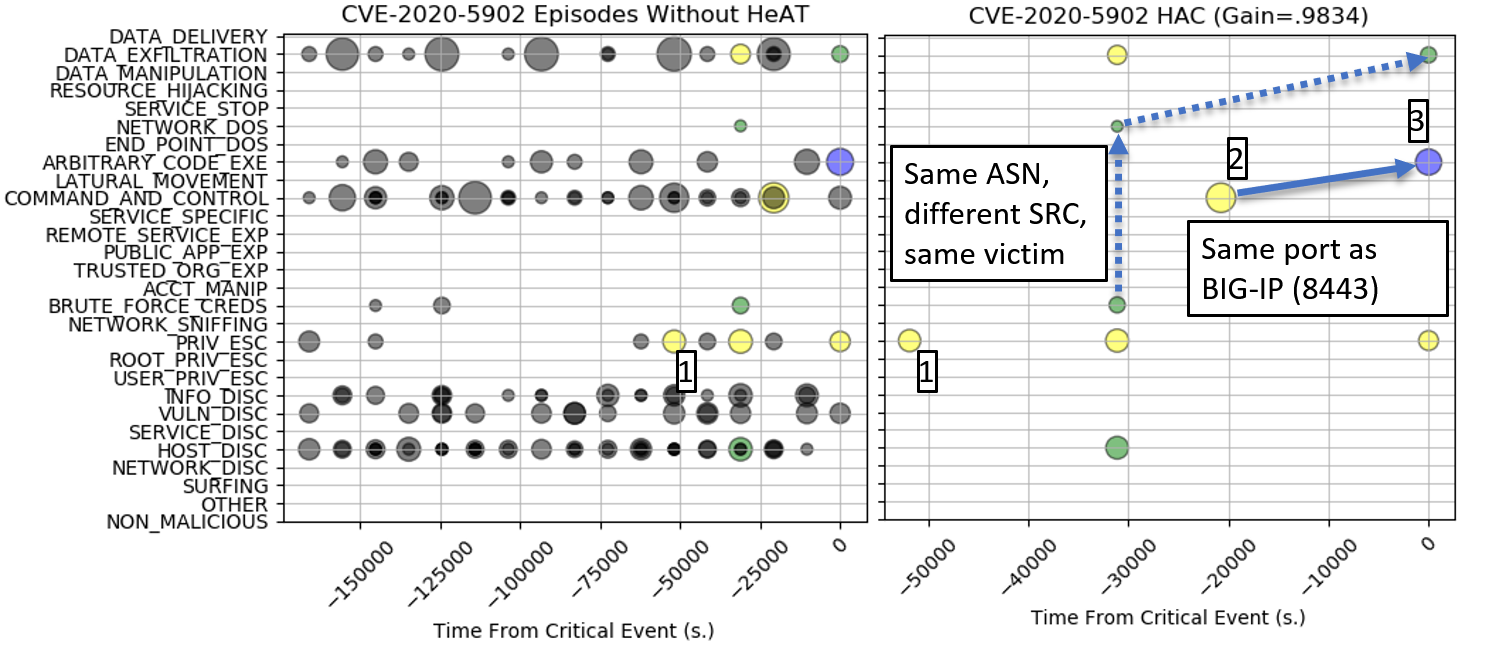}
    \caption{HAC's extracted for CVE-2020-5902 with (right) and without (left) using HeAT. The HAC extracted by HeAT substantially reduces irrelevant alert episodes and reveals concerning activities leading to the critical exploitation.}
    \label{fig:soc_ex1}
\end{figure*}

We now turn to the alert associated with CVE-2020-5902, which describes a remote code execution (RCE) vulnerability on F5's enterprise traffic manager, load balancer, and DNS.
This is a particularly critical signature; if successful, the adversary will have nearly full control over the flow of traffic within the network and could redirect traffic wherever they desire.  
On the right of Figure \ref{fig:soc_ex1}, we show the HAC extracted by HeAT with the highest HeAT Gain and AIS coverage with alert episodes occurring within 48-hours prior to the critical event. To give a visual comparison, we show an HAC without using HeAT on the left of Figure \ref{fig:soc_ex1} -- there are many more alert episodes within the 48-hr time frame and turns out to be irrelevant.

This example exemplifies the usefulness of HeAT where it reduces significantly the irrelevant alerts while keeping critical alert episodes that actually needs the analyst's attention to perform in-depth analysis on how the multistage attack including lateral movements transpired. Given this HAC, an analyst is likely to be able to make a timely assessment as follows (following the annotations on the right of Figure \ref{fig:soc_ex1}):

\begin{enumerate}
    \item Evidence of root-privilege escalation on the victim.
    \item C2 activity on the same source, victim and port number as the IoC.  Unique to this HAC.
    \item The IoC episode with the same destination port as the C2, likely compromised.  A separate exfiltration  was found from the same ASN occurring simultaneously.
    \item (Recommendation) Investigate the C2 behavior and how the initial privilege escalation happened.
\end{enumerate}

\subsection{Transferability of HeAT Model} \label{sec:transferrability}
Given the promising case-study shown above, we attempt to conduct a broader set of experiments to determine how and whether adding a small set of additionally labeled episodes from the SOC data can help to reduce the HeAT Coherency. 
Due to the vast difference in alert volume, network architecture, and attacker behavior between the CPTC and the SOC data, we assume that additional analyst's assessment for the SOC data will help improve extracting HAC's tailored for the SOC network.
In the following experiments, we further add to the training data by assessing two IoC instances associated with `\emph{ET EXPLOIT Possible ETERNALBLUE Probe MS17-010}' from the SOC data.
This signature signifies actions that allow for arbitrary code to be executed through a SMB-share, typically leading to the installation of ransomware.
HeAT values were assigned for 125 episodes prior to these two instances and used to include for HeAT model training.
Note that the 125 episode labeling is a small fraction of the 1,400+ observations used for the initial training with CPTC data only and it is not unreasonable to ask an analyst to do so.

\begin{figure}[h]
    \centering
    \includegraphics[width=8cm,height=5cm]{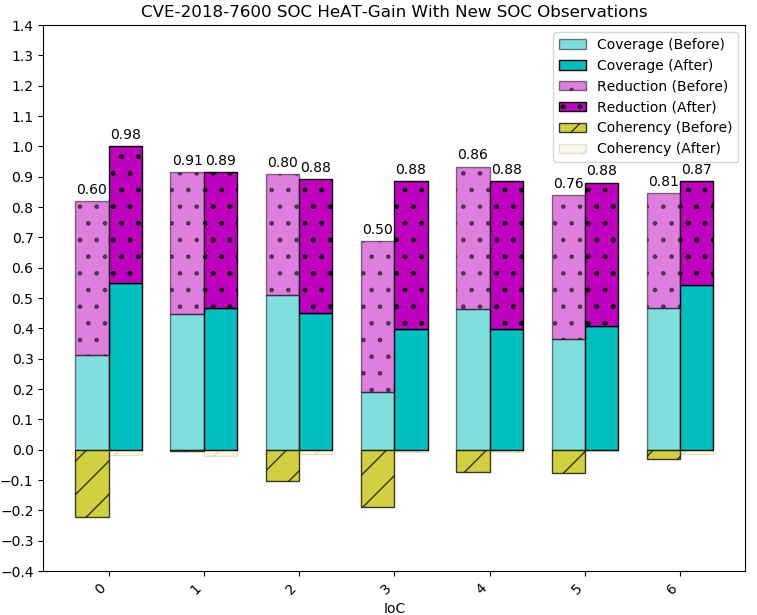}
    \caption{The HeAT Gains of HAC's extracted by HeAT for the 7 IoC's (CVE-2018-7600) with (right bar) and without (left bar) using the additional SOC assessments for training.}
    \label{fig:soc_trans}
\end{figure}

Figure \ref{fig:soc_trans} shows the HeAT-Gain of the seven HAC's for the IoC's associated with a distinct Suricata signature `Drupalgeddon2 CVE-2018-7600' 
before and after the new SOC assessments are added to train the HeAT model.
the HeAT Gains of the HAC's with respect to the . 
It can be clearly seen that adding additional SOC observations reduces the coherency to near-zero across all instances.  
We also find that in some cases the CPTC-only model performs well while adding new assessments does not hurt the performance.

A deeper analysis of these experiments lead to a contextual question for the machine learned model: \textit{What attack campaign characteristics are getting transferred and what are improved with the new assessments?}
Recall that in our CPTC experiments that the model was biased to episodes that had the same source and target IP addresses of the IoC.
We find that within SOC data it is common to observe a single episode with multiple sources and many targets.
This is a result of us finding that filtering by source-ASN was much more appropriate to represent and aggregate the vast volume of SOC alerts.
Remember that our network-agnostic features evaluate the \textit{ratio} of common IP addresses.
We found in cases where there was a large discrepancy in the HeAT coherency, such as IoC 0 and 3 in Figure \ref{fig:soc_trans}. The CPTC-only model would apply high heat to episodes where there was almost an exact match between the source and target IP.
With the additional SOC observations, HeAT was able to bring up the HeAT-value for cases where the true victim IP address was one of many that were targeted within that episode.
This would not be unusual since with a network of this scale, many adversaries may target many victims in the hopes of gaining access to just one.
This result demonstrates that a small number of additional ASN-based observations enable our IP-based observations in CPTC to be accurately applied and find attack campaigns within network data multiple orders of magnitude larger. 

IoC's 1, 4, and 6 demonstrate cases where only the CPTC observations are enough to accurately capture attack campaigns, with IoC 4 actually reporting a higher gain than if we supplied additional data.
IoC 4 reports a higher gain using CPTC-only due to applying heat to episodes that are identified as a false positive attack response that triggers a scanning type alert at the same time as the IoC.
Our additional data has made those types of alerts low heat and excluded from the HAC.
This led us to finding another bias, the CPTC data model is biased towards applying HeAT of episodes with close time proximity to the IoC and applying high HeAT to benign signatures.

We also investigated the behavior of HeAT for all unique signatures, including non-malicious signatures, as we expect lesser critical signatures to have little HeAT.
The top-3 signatures with the highest gain given the CPTC-only observations consisted of: 1) \textit{ET WEB\_SERVER Suspicious Chmod Usage in URI (Inbound)} ($\Delta$=1.04), 2) \textit{ET WEB\_SERVER 401TRG Generic Webshell Request - POST with wget in body} ($\Delta$=1.01), and 3) \textit{ET WEB\_SERVER WebShell Generic - wget http - POST} ($\Delta$=.99).
From their descriptions, it is clear that these signatures are not the result/end of a critical and impactful attack campaign but instead due to CPTC not accurately reflecting all characteristics of a SOC data-set.
The additional SOC data pushed down the HAC's associated with these non-malicious signatures to the bottom-5 gains and brought up the two CVE signatures. 
While our network-agnostic features on their own can provide reasonable results, it is also clear that the additional fine-tuning from the SOC data-set provides more accurate severity assessments with HeAT-gain.   
We demonstrate that despite the vast differences between the CPTC and real-world SOC operation, HeAT was able to transfer many meaningful characteristics to extract effectively unseen attack campaigns.

\section{Conclusion}
\label{sec:conclusion}
HeAT and its ability to capture the analysts opinion of events likely to contribute to an attack campaign and apply that knowledge to discover attack campaigns in other networks is a unique approach to combat the reliance on large labelled data sets.  
In a field where there is significant privacy concerns with the simply the access and usage of unlabelled data sets containing adversarial behavior, we find it advantageous that competition data like CPTC can indeed be used to find other attack campaigns within the SOC data set with minimal extra effort needed.
We envision the scenario where HeAT can provide initial professionally labelled data that is dense with actual attack campaign information and ask the end user to provide additional data to fine-tune the model for their network.  
This eliminates the privacy concerns when using observations from other networks as our episode features never contain any specific information either network while maintaining critical attack campaign indicators.
HeAT is unique in the sense that few works demonstrate their ability to provide meaningful results in both highly controlled scenarios and in a real-world scenario where little is known.
For the foreseeable future, triaging networks to assess attack campaigns is only going to get more difficult and time consuming to the point that manual triaging becomes infeasible. As evidenced by efforts in the private sector, automated triaging with AI/ML will be a necessary component to all SoC operations.
We have shown that HeAT is a viable solution that requires minimal expertise and analyst' effort. 
Future development of HeAT includes integrating with Zeek logs and phishing email detection and developed as a plug-in with Splunk or other SIEMs.



\bibliographystyle{IEEEtran}
\bibliography{moskal}

\end{document}